\documentclass[12pt]{article}
\usepackage[utf8]{inputenc}

\usepackage{amssymb}
\usepackage{amsmath}

\usepackage{hyperref}
\usepackage{lineno}

\usepackage[normalem]{ulem}
\usepackage{color}
\usepackage{xspace}

\usepackage{authblk}

\usepackage{graphicx}

\usepackage[comma,square,numbers,sort&compress]{natbib}

\newcommand{\snn}{\ensuremath{ \sqrt{s_{\rm NN} } } }

\newcommand{\PbPb}{\ensuremath{{\text{Pb--Pb}}}}

\newcommand{\TCF}{\ensuremath{T_{\rm CF}}}

\newcommand{\muB}{\ensuremath{\mu_{\rm B}} }
\newcommand{\kB}{\ensuremath{k_{\rm B}} }
\newcommand{\chisq}{\ensuremath{\chi^{2}/N_{\rm dof}} }

\title{Influence of modified light-flavor hadron spectra on particle yields in the statistical hadronization model}

  \author[1]{A.~Andronic}
  \author[2,3,4]{P.~Braun-Munzinger}
  \author[3]{D.~G\"{u}nd\"{u}z}
  \author[3]{Y.~Kirchhoff}
  \author[3]{M.~K.~K{\"o}hler}
  \author[2,3]{J.~Stachel}
  \author[5]{M.~Winn}

  \affil[1]{ Institut f\"{u}r Kernphysik, Westf\"{a}lische Wilhelms-Universit\"{a}t M\"{u}nster, M\"{u}nster, Germany}
  \affil[2]{Research Division and ExtreMe Matter Institute EMMI, GSI Helmholtzzentrum f\"{u}r Schwerionenforschung GmbH, Darmstadt, Germany}
  \affil[3]{Physikalisches Institut, Ruprecht-Karls-Universit\"{a}t Heidelberg, Heidelberg, Germany}
  \affil[4]{Institute of Particle Physics and Key Laboratory of Quark and Lepton Physics (MOE), Central China Normal University, Wuhan 430079, China}
  \affil[5]{D\'{e}partment de Physique Nucl\'{e}aire (DPhN), Paris-Saclay Centre d'Etudes de Saclay (CEA), IRFU, Saclay, France}

\date{\today}

\begin{document}
\maketitle

\begin{abstract}
    Hadron production in relativistic nuclear collisions is well
    described in the framework of the Statistical Hadronization Model
    (SHM). We investigate the influence on SHM predictions of hadron
    mass spectra for light-flavor baryons and mesons modified by the
    addition of about 500 new states as predicted by lattice QCD and a
    relativistic quark model. The deterioration of the resulting
    thermodynamic fit quality obtained for $\PbPb$ collision data at
    $\snn = 2.76$~TeV suggests that the additional states are not
    suited to be naively used since also interactions among the states
    as well as non-resonant contributions need to be considered in the
    SHM approach. Incorporating these effects via the pion nucleon
    interaction determined from measured phase shifts leads again to
    excellent reproduction of the experimental data. This is a strong
    indication that at least the additional nucleon excited states
    cannot be understood and used as independent resonances.
\end{abstract}

\section{Introduction}

The wealth and precision of hadron-yield measurements in heavy-ion
collisions at the LHC provide new insights into the hadronization
process and the preceding collision stages. Within the framework of
the statistical hadronization model~(SHM) it can be shown that, under
the assumption of a rapid chemical freeze-out of a thermalized medium,
particle abundances are described by statistical weights with only a
small set of parameters: the chemical freeze-out temperature~$\TCF$,
the baryo-chemical potential $\muB$, and the fireball
volume~$V$~\cite{Andronic:2005yp,Becattini:2005xt,Cleymans:2006qe,Andronic:2008gu,Floris:2014pta,Andronic:2017pug,Vovchenko:2019kes}. The
particle species considered comprise the lightest and most abundantly
produced hadrons such as pions as well as more complex and even
loosely bound objects, such as (anti-)(hyper-)nuclei and all measured
yields can be described with very good
accuracy~\cite{Andronic:2017pug,Braun-Munzinger:2018hat}. The approach
can also be extended to the heavy-flavor
sector~\cite{BraunMunzinger:2000px,Andronic:2017pug,Andronic:2019wva}.

The SHM connects measurements of particle yields to the phase diagram
of strongly interacting matter, which is resulting from Quantum-Chromo
Dynamics~(QCD). In particular, hadron production is governed by the
QCD phase transition at a pseudo-critical temperature $T_c$ between
the Quark-Gluon Plasma~(QGP), a state with quarks and gluons as
degrees of freedom, and the confined phase where hadrons are the
degrees of freedom~\cite{BraunMunzinger:2008tz}. The equations of QCD
cannot be solved perturbatively for values of the strong coupling
constant appropriate for typical particle distances and momentum
transfers in the QGP. So far this can be done only numerically by
putting the theory on a discretized space-time lattice in a method
called lattice QCD~(LQCD), see e.g. reference~\cite{Ratti:2018ksb} for
a recent review. An alternative is modelling QCD with effective field
theories. The statistical operator emerging from LQCD studies is in
fact in very good agreement with that based on the hadron resonance
gas approach and used in the analysis of hadron yields in the
framework of the SHM, supporting the conjecture of quark-hadron
duality close to the QCD phase boundary~\cite{Andronic:2017pug}.

LQCD at zero temperature, as well as quark models with two or three
constituent quarks, can be used for a prediction of baryon and meson
mass spectra. These calculations can be confronted with experimental
spectroscopic data to shed light on the degrees of freedom in the QCD
mass spectrum.

In the SHM, the mass spectrum of hadronic states is needed to build
the hadronic partition function. In previous investigations
(\cite{Andronic:2017pug} and references therein), the hadron mass
spectrum was based entirely on experimental data as compiled by the
Particle Data Group (PDG)~\cite{PDG:2018}. The resulting mass spectrum
is extraordinarily rich~\cite{Lutz:2015ejy} and the density of states
grows approximately exponentially up to the mass region of about 1.8
GeV, from where on the increase starts to flatten.  This exponential
growth was in fact predicted long ago~\cite{Hagedorn:1965st} and led
to the concept of a limiting temperature for hadronic
matter. Incidentally, the value of $\TCF$ obtained from SHM analyses
is somewhat below this limiting temperature and, as demonstrated
in~\cite{Andronic:2017pug}, can be identified with the pseudo-critical
transition temperature $T_c$ between QGP and hadronic matter.

Above about 2 GeV, the states become very dense, the hadronic
resonances start to strongly overlap and experimental identification
becomes very difficult. The flattening of the experimentally known
density of states at large mass prompted a
study~\cite{Andronic:2008gu} to what extent the SHM predictions and
fits of data would be influenced by missing states. Assuming a
continued Hagedorn-type growth, an effect on the pion yield of up to 9
\% was found. The experimental considerations just mentioned as well
as comparison to LQCD predictions indicate that even in the mass range
below $1.8$~GeV the mass spectrum could be incomplete
(\cite{Lo:2015cca} and references there). This has led to
attempts~\cite{Bazavov:2014xya}, in the context of reconciling LQCD
data on strangeness fluctuations with the HRG, to supplement the
experimental data with theory input for the hadron mass spectrum.
This input came, e.g., from LQCD~\cite{Edwards:2011jj,Edwards:2012fx}
as well as the relativistic quark
model~\cite{Ebert:2009ub,Capstick:1986bm}.  A comparison of SHM
predictions based on such a modified mass spectrum with data has not
been made yet at the level of the first moments, i.e. the hadron
yields. We note that, fortunately, the sensitivity to higher masses of
SHM predictions for hadron yields is strongly reduced because the
experimentally determined chemical freeze-out temperature $\TCF$ is
less than 160 MeV~\cite{Andronic:2017pug} and, hence, thermal
population of hadrons with masses larger than 1.5 GeV is strongly
Boltzmann suppressed. Nevertheless, since in these theoretical
investigations many additional states are proposed, and some of them
with large spin degeneracies, one should study their possible
influence on hadron production in relativistic nuclear
collisions. This is the aim of the present paper.

We conclude this introductory section with a remark that is becoming
increasingly relevant as the SHM predictions are now tested with the
LHC data at precisions well below the 10\% level over a data range of
more than 9 orders of magnitude. The case in point is the assumption,
in the current formulation of the SHM, that the hadron mass spectrum
can be obtained from the PDG compilation~\cite{PDG:2018} of vacuum
(zero temperature) experimental hadron masses and branching
ratios. The successes reported in
~\cite{Andronic:2017pug,Braun-Munzinger:2018hat} indeed lend strong
experimental support to this assumption. This is, however, in stark
contrast to predictions from LQCD~\cite{Aarts:2017iai,Aarts:2017rrl}
and refs. there that hadron masses should change for small
baryo-chemical potential as the temperature reaches values in the
assumed cross over phase transition region. Yet, despite many
experimental efforts, no convincing experimental evidence of any
in-medium hadron mass change (apart from collisional broadening of the
$\rho$ meson, ~\cite{Rapp:1999ej}) has been reported. The possibility
of modified hadron masses in the SHM was analyzed
in~\cite{Florkowski:1999pz,Michalec:2001qf}. There it was demonstrated
that, to keep a good description of experimental data, scaling of
masses necessitates a similar scaling of the chemical freeze-out
temperature. This, however, is inconsistent with dropping masses as
the temperature increases towards the (pseudo-)critical value.

Recently it was argued~\cite{Bluhm:2020rha} that, at LHC energies, the
chiral condensate effectively decouples from the thermal fireball, and
relaxes quickly to the vacuum value, possibly leading to vacuum masses
as employed in the SHM. However, in thermal equilibrium, such effects
should be contained in the LQCD analysis.

A way out, however conflicting with LQCD predictions~\cite{Aoki:2006we}, could be that the phase transition is first order after all, but leads, directly below $T_c$, to a non-equilibrium state  with vacuum masses that cannot be described under the complete equilibrium conditions of LQCD. Clearly, the absence of any firm experimental information on the order of the QCD phase transition is not satisfactory  and calls for further experimental and theoretical investigations. 

\section{Model description}

The description of the model given here is restricted to its basic
features and focuses on the details of the hadronic states implemented
in the SHM. A more detailed description of the model and the
underlying assumptions can be found in~\cite{Andronic:2017pug} and
references there.

The basic quantity for thermodynamic computations is the partition
function $Z(T,\mu,V)$. For a hadron species $i$ in the grand-canonical
ensemble it can be written for an ideal (non-interacting) quantum gas
in the following form\footnote{Natural units, $\hbar = c = \kB = 1$,
  are used throughout this document.}:
\begin{align}
\ln{Z_i}  &= \frac{V g_i}{2 \pi^2} \int^\infty_0 \pm p^2 \text{d} p \ln \left( 1 \pm \exp{  [ -(E_i-\mu_i)/T ] } \right),
\end{align}
where $g_i = 2(J_i + 1)$ is the degeneracy factor for the total
angular momentum $J$, $p$ is the particle momentum, and $E_i =
\sqrt{p^2 + m_i^2}$ it's total energy. The sign $(+)$ is for fermions
and $(-)$ for bosons. The chemical potential is given as $\mu_i = \muB
\cdot B_i + \mu_S \cdot S_i + \mu_{I3} \cdot I_{3i} + \mu_C \cdot
C_i$, where $B$, $S$, $I_3$, and $C$ are the baryon, strangeness,
$3$-component of the isospin and charm quantum numbers.

From this partition function, the particle densities $n_i$ and all
thermodynamic quantities can be derived. In particular,
\begin{align}
n_i(T,\mu_i) &= N_i/V = -\frac{T}{V} \left( \frac{\partial \ln Z_i}{\partial \mu_i } \right)_{V,T} \\
&= \frac{g_i}{2 \pi^2} \int^\infty_0 \frac{p^2 \text{d} p}{\exp{ [(E_i-\mu_i)/T] } \pm 1}.
\end{align}

In addition, an eigenvolume correction may be applied in order to
model repulsive interactions between different hadrons, as studied in
detail in~\cite{Andronic:2012ut}. This leads to an iterative procedure
for the consistent determination of the thermodynamic quantities. In
particular, the particle densities are corrected by a common factor
reducing all particle densities while yield ratios are, in general,
unchanged. The impact of modeling interactions beyond this ansatz via
the $S$-matrix approach~\cite{Lo:2017lym} is discussed later in the
context of the description of hadron yields with the modified particle
spectra.

The total partition function of the model for $ T \lesssim \TCF$ can
be written as a sum of mesonic and baryonic contributions:
\begin{equation}
  \begin{aligned}
    \ln Z(T, \mu, V) = & \sum_{i \in \text{ mesons}} \ln Z_i(T,\mu_{I3},\mu_S,\mu_C, V) + \\
    & \sum_{i \in \text{ baryons}} \ln Z_i(T,\mu_B,\mu_{I3},\mu_S,\mu_C, V).
  \end{aligned}
  \label{eq:partition}
\end{equation}

For comparison with experimental data, the particle yields are
computed from the thermal densities $n_i$ taking into account all
strong decays. Starting from the thermal densities, decays are
propagated via a decay matrix which gives statistical weights to the
corresponding daughter particles incorporating the individual
branching ratios. If the branching ratios for decays of a particle
with mass $m$ are not known, as is the case for particles introduced
in this study on theoretical grounds but not established
experimentally yet, the branching ratios are estimated based on the
known systematics of the 500+ established hadrons and considering the
decay phase space. For a hypothetical strong decay of the particle
with mass $m$ into a given final state with two daughter particles and
a decay momentum $|\mathbf{p}_1|$ the decay rate into the solid angle
$\text{d} \Omega$ can be written as~\cite{PDG:2018}
\begin{equation}
    \text{d}\Gamma = \frac{1}{32 \pi ^2} |\mathcal{M}|^2 \frac{|\mathbf{p}_1|}{m^2} \text{d}\Omega,
    \label{eq:decaymatrix}
\end{equation}
where $\mathcal{M}$ is the Lorentz-invariant matrix element. Assuming
that for competing strong decays of the hypothetical particle with
mass $m$ the genuine matrix element is typical for the strong decay
and not dif\-fe\-rent for dif\-fe\-rent decay channels, we only
consider the phase space factor to determine the branching ratios into
different final states as given by the decay momenta. The decay
channels considered are obtained from the known decays of hadrons with
the same quantum numbers and considering the opening up of new allowed
decay channels with increasing decay momentum. Taking into account
only 2-body decays implies that we effectively treat all newly opening
decays of states with unmeasured branching resonantly. A newly
introduced $\Delta$ resonance will decay partly via a lower-lying
$\Delta$ resonance as soon as energetically possible in addition to
the direct decay to nucleon plus pion, which will in turn decay into
nucleon plus pion thus effectively increasing the number of pions in
the final state as phase space opens up. This is exactly what is
measured for the known excited nucleon resonances. Similar arguments
apply to the meson sector with intermediate $\rho$ or $\eta$ or $K^*$
modes.

The particle list in the standard version of the SHM contains $555$
established hadrons as given by the PDG~\cite{PDG:2018}. This list of
particles consists of $223$ mesons ($123$ non-strange, $32$ strange,
$40$ with charm, and $28$ with beauty) and $270$ baryons ($104$
non-strange, $96$ strange, $56$ with charm, and $14$ with beauty). It
includes, in addition, $62$ nuclei, hyper-nuclei and
anti-\-(hyper-)\-nuclei.

In the following, we discuss the impact of extending the hadron mass
spectrum with particles expected from LQCD and the relativistic quark
model.

\begin{figure}[h]
    \centering
    \includegraphics[width=1.0\textwidth]{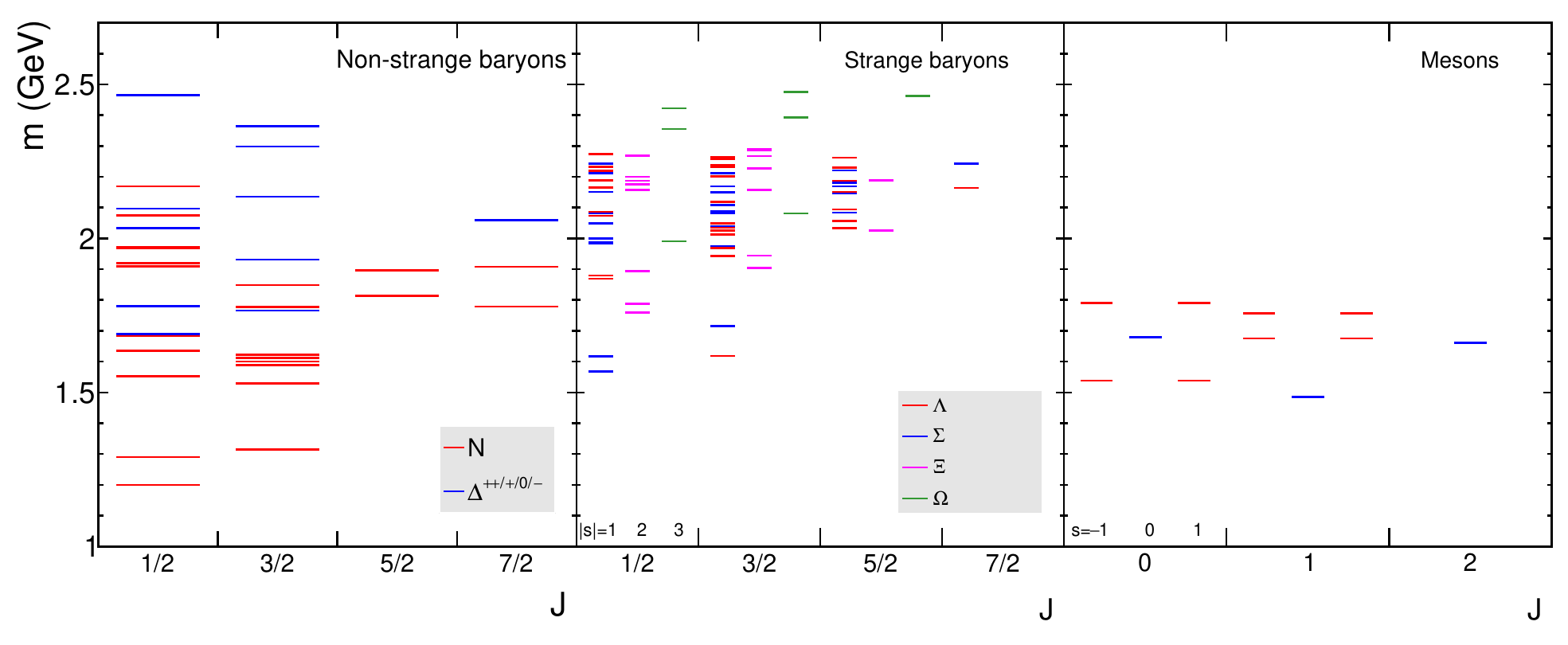}
    \caption{A synopsis of the additional states in terms of mass and
      total angular momentum for non-strange baryons (left panel),
      strange baryons (middle panel) and mesons (right panel.}
    \label{fig:plot_AllStates}
\end{figure}

\section{Extension of the particle spectrum}
\label{sec:extension}

Using theoretical input, the particle spectrum used in the SHM in this
study is extended in 3 steps with respect to the original list based
on experimentally established particles as given by the
PDG~\cite{PDG:2018}:
\begin{enumerate}

\item In a spectrum labelled 'udB' non-strange baryon states $N^*$
  with isospin I=1/2 and $\Delta^*$ with isospin I=3/2 are added as
  predicted by a LQCD calculation~\cite{Edwards:2011jj}. Up to a mass
  of about 2.2 and 2.5 GeV, respectively, 23 $N^*$ doublets and 11
  $\Delta^*$ quadruplets and the corresponding anti-baryons with total
  angular momenta between 1/2 and 7/2 have been added. The LQCD
  calculations have been performed with unphysically massive light
  quarks resulting in pion masses between 396 and 524 MeV. Guided by
  the baryon mass evolution with decreasing light quark mass for the
  calculations shown in ~\cite{Edwards:2011jj}, we scale down the
  masses of the additional $N^*$ and $\Delta^*$ states from the LQCD
  results for $m_{\pi}$ = 396 MeV by a factor obtained comparing the
  masses of the already known non-strange baryon states. The scale
  factors are 0.71 and 0.76 for the $N^*$ and $\Delta^*$ states. Since
  several of the states are very closely spaced in mass, for
  convenience we group these states together in the code and adjust
  the corresponding degeneracy factors accordingly. This extension
  increases the number of non-strange baryons states from 104 to
  284. The added states are visualized in the left panel of
  Fig.~\ref{fig:plot_AllStates}.

\item In a next step, in the spectrum labelled 'udsB', strange baryons
  are added as predicted in LQCD calculations by the same
  authors~\cite{Edwards:2012fx}. Also in these calculations, the light
  quark masses are unphysically high resulting in a pion mass $m_{\pi}
  = 391$~MeV. In total 32 $\Lambda$ isospin singlets, 26 $\Sigma$
  isospin triplets, 17 $\Xi$ isospin doublets and 7 $\Omega$ states as
  well as the according anti-baryons have been added. Following the
  same approach as chosen for the non-strange baryons, the mass
  spectra are scaled to the known states for the according multiplets,
  resulting in scaling factors 0.85, 0.88, 0.95 and 1.0 for the
  $\Lambda$, $\Sigma$, $\Xi$, and $\Omega$ states, respectively. Due
  to the decreasing number of light valence quarks for the different
  isospin multiplets, it is plausible that scale factors approaching 1
  are found with increasing strangeness. After scaling, an upper mass
  limit for the added baryons was chosen of $2.3$~GeV for $\Lambda$,
  $\Sigma$, and $\Xi$ states and $2.5$~GeV for $\Omega$ states,
  resulting in the numbers of states given. In total the spectrum now
  contains 398 instead of originally 96 strange baryon states. The
  added states are shown in the middle panel of
  Fig.~\ref{fig:plot_AllStates}.

\item Finally, additional meson states are added. Since there is no
  LQCD result corresponding to the baryon
  predictions~\cite{Edwards:2011jj,Edwards:2012fx}, the spectrum
  'udsBM' with added strange and non-strange meson states is based on
  a relativistic quark model~\cite{Ebert:2009ub}. In total 2
  non-strange triplets, 2 non-strange singlets and 4 $K^*$ doublets
  and their anti-doublets are added. The mesons are added up to a mass
  limit of $1.7$~GeV and $1.8$~GeV for non-strange and strange
  multiplets, respectively. The increase in the total number of mesons
  is moderate, just about 10 \%. The added states are shown in the
  right panel of Fig.~\ref{fig:plot_AllStates}.

\end{enumerate}

The decay matrix for the additional states is deduced from
experimentally known states with the same quantum numbers as described
above and considering a mass dependence due to the growing phase space
with increasing mass as given in eq.~\ref{eq:decaymatrix}.

\begin{figure}[htb]
    \centering
    \includegraphics[width=0.55\textwidth]{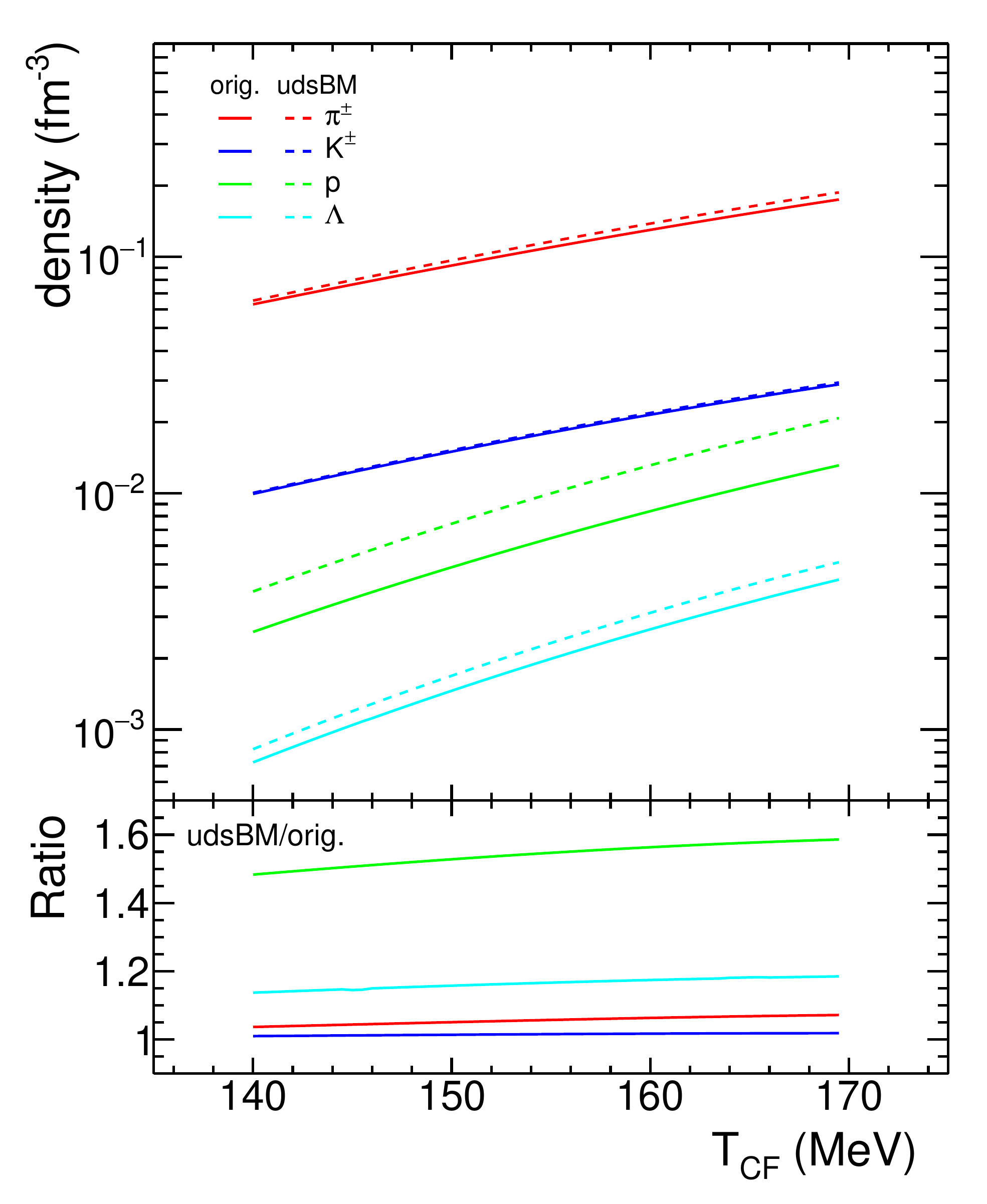}
    \caption{Temperature dependence of the thermal densities after
      strong decays for pions, kaons, protons, and $\Lambda$ hyperons
      for the original (PDG) hadron spectrum and after the inclusion
      of the additional states (upper panel). The corresponding ratios
      are shown in the lower panel. The absolute values of the
      densities are for a calculation with an excluded volume
      parameter of 0.3~fm, to obtain values without excluded volume
      one needs to multiply by a factor 1.34. }
    \label{fig:plot_density}
\end{figure}

The added states decay strongly into nucleons, the lightest strange
baryons of a given strangeness, pions, and kaons. Correspondingly, the
densities of those baryons and mesons change when new states are added
and the change is temperature dependent, since with increasing
temperature the added states are populated increasingly. The
temperature dependence of the thermal densities for pions, kaons,
protons and $\Lambda$ hyperons after strong decays (and
electromagnetic decay of the $\Sigma^0$) is shown in
Fig.~\ref{fig:plot_density} for the original (PDG) hadron spectrum and
after the inclusion of all additional states, corresponding to
spectrum 'udsBM'.

While for the temperature range shown in Fig.~\ref{fig:plot_density}
the effect on the pion density is very small (sub 10 \%) and even less
for the kaons, for the protons a significant increase in density of 50
to 60 \% is observed due to the additional states. The large increase
is due to the addition of some very low mass $N^*$ resonances
resulting from the LQCD calculation. They appear well below the Roper
resonance ($N^*(1440)$). For the $\Lambda$ hyperons, the increase is
smaller but not negligible with 15 to 20 \%.

\section{Resulting thermal fits}

\begin{figure}[htb]
    \centering
    \includegraphics[width=1.0\textwidth]{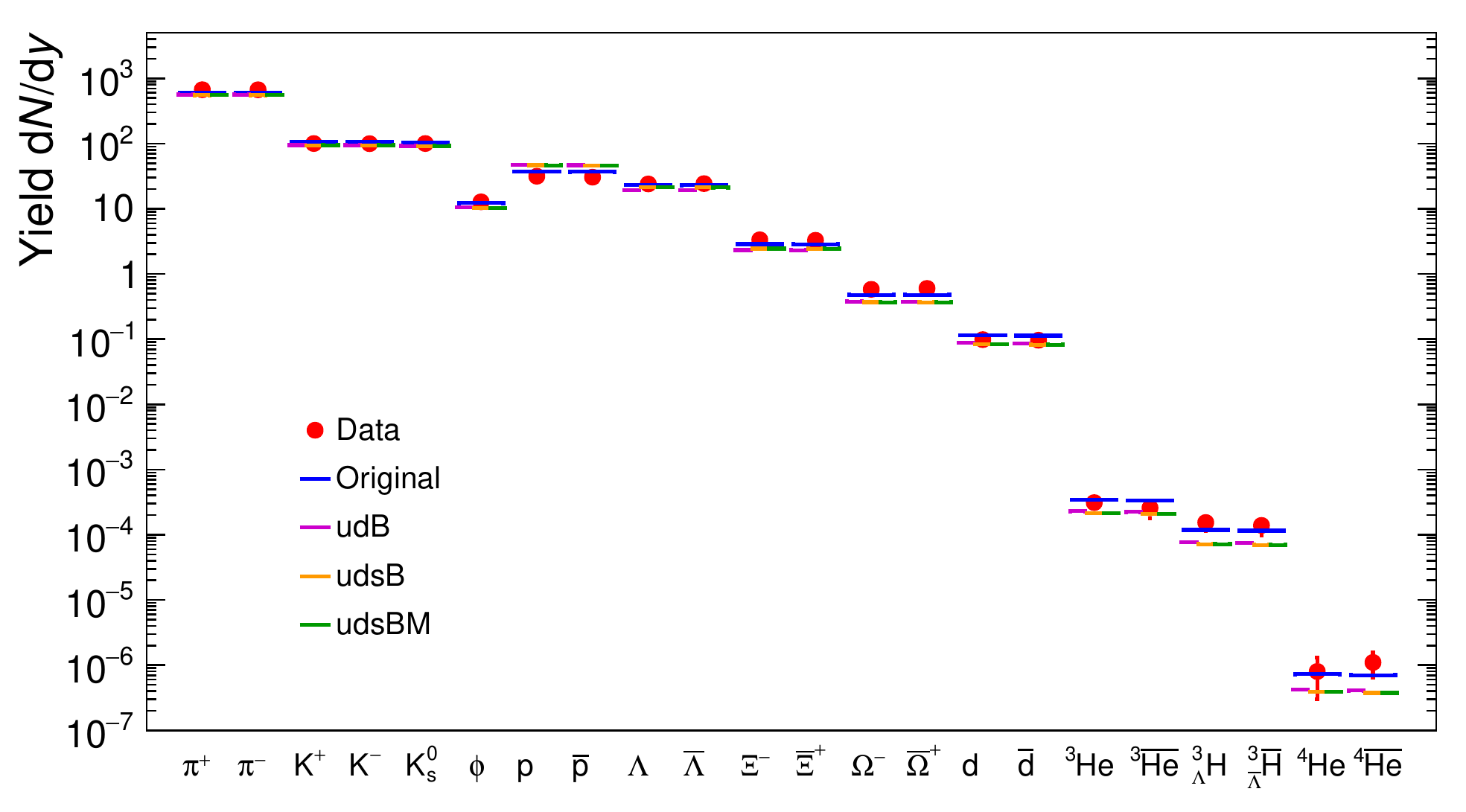}
    \caption{Comparison of particle yields measured by ALICE at
      mid-rapidity (references see text) to the fit using the original
      (PDG) spectrum and model fits using the 3 modified hadron
      spectra. }
    \label{fig:plot_yieldAll}
\end{figure}

In this section we confront the SHM predictions using the modified
hadron spectra to experimental hadron yields from the LHC. In analogy
to the thermal fit discussed in~\cite{Andronic:2005yp}, we perform
fits based on a minimization of the $\chi^2$ value between
experimental hadron yields at mid-rapidity and SHM results varying the
3 free parameters. Here we use the ALICE data for Pb-Pb collisions at
a collision energy per colliding nucleon pair of $\snn =
2.76$~TeV. Experimental results for $22$ particle species are used in
the
fits~\cite{Abelev:2013vea,Abelev:2013xaa,ABELEV:2013zaa,Abelev:2014uua,Adam:2015yta,Adam:2015vda,Acharya:2017bso}. Effectively,
there is only one free fit parameter to reproduce the yields of
identified hadrons, the chemical freeze-out temperature $\TCF$,
because for this collision system the baryon chemical potential is
consistent with zero in all fits and the volume is fixed already by
the overall charged particle multiplicity. Fit results for the
original and the $3$ modified hadron spectra are shown in
Fig.~\ref{fig:plot_yieldAll} together with the experimental
data. Using the original (PDG) hadron spectrum leads to results
identical to the calculations reported in~\cite{Andronic:2017pug}.

\begin{figure}[htb]
    \centering
    \includegraphics[width=1.0\textwidth]{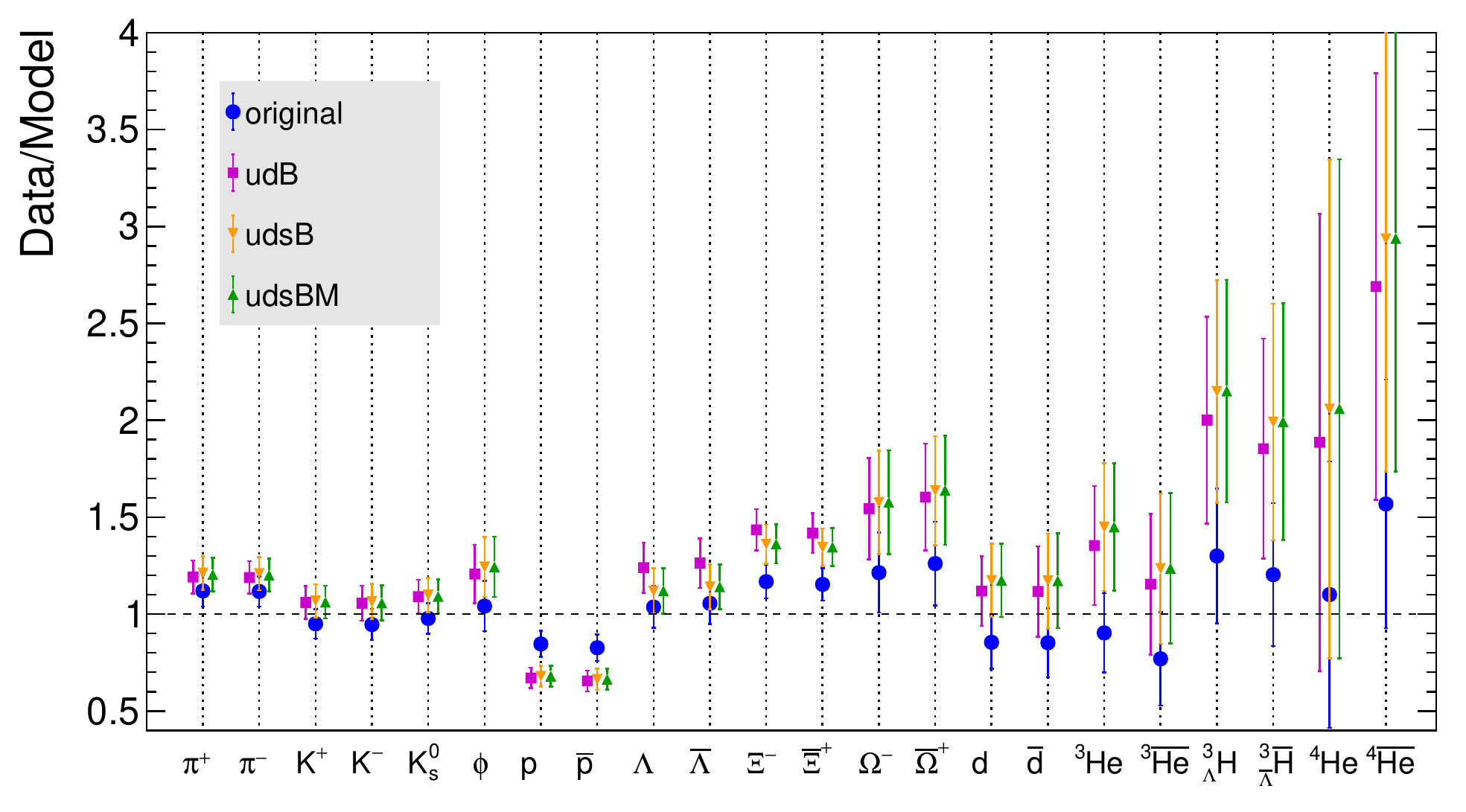}
    \caption{Ratio of ALICE experimental data over model fit for
      yields of all 22 fitted particle species for the original and
      the 3 modified hadron spectra. }
    \label{fig:plotratio_yieldAll}
\end{figure}

To allow a better visual comparison between the experimental data and
the different thermal fits, the corresponding ratios for all fits are
shown in Fig.~\ref{fig:plotratio_yieldAll}. The experimental data
points are well described in case of the original PDG particle
list. In contrast, a significant and systematic deviation between data
and calculations is observed when using the modified hadron mass
spectra.

Both figures show clear modifications for all particle species between
the original fit and the fits with modified hadron spectra. The $3$
fits with additional states group together for most particle species,
with the exception of the strange baryon sector, where the fit using
the spectrum 'udB' shows the largest discrepancy to the original fit
result, while using the spectra 'udsB' and 'udsBM' represents an
improvement, in particular for the $\Lambda$. The key to this
observation is already visible in Fig.~\ref{fig:plot_density}. Adding
the additional states \textsl{increases} the proton yield in the
model. However, already using the original PDG hadron spectrum results
in a proton yield high by about 15 \% compared to experimental
data. Incidentally, this deviation from the experimental data by 2.6
standard deviations is the only significant discrepancy between data
and model with the original PDG spectrum (and in the absence of the
S-matrix correction, see below). Therefore it is immediately clear
that a further increase in the model proton yield can only lead to
significant deterioration since the proton yield is measured with high
precision. In line with this observation, with the additional states
the proton discrepancy grows and drives the fit to a smaller
temperature to mitigate in part the detrimental effect of an increased
proton yield, still yielding a larger than 30 \% deviation for the
proton yield. The drop in temperature is 3.5 MeV for spectrum 'udB'
and 4 MeV for spectra 'udsB' and 'udsBM'. Due to this drop in
temperature, the yields for all other hadrons start to differ
significantly, as visible in Fig.~\ref{fig:plotratio_yieldAll}. With
the temperature in the model fit reduced, all yields other than the
protons are under-predicted more or less significantly.

\begin{figure}[htb]
    \centering
    \includegraphics[width=.6\textwidth]{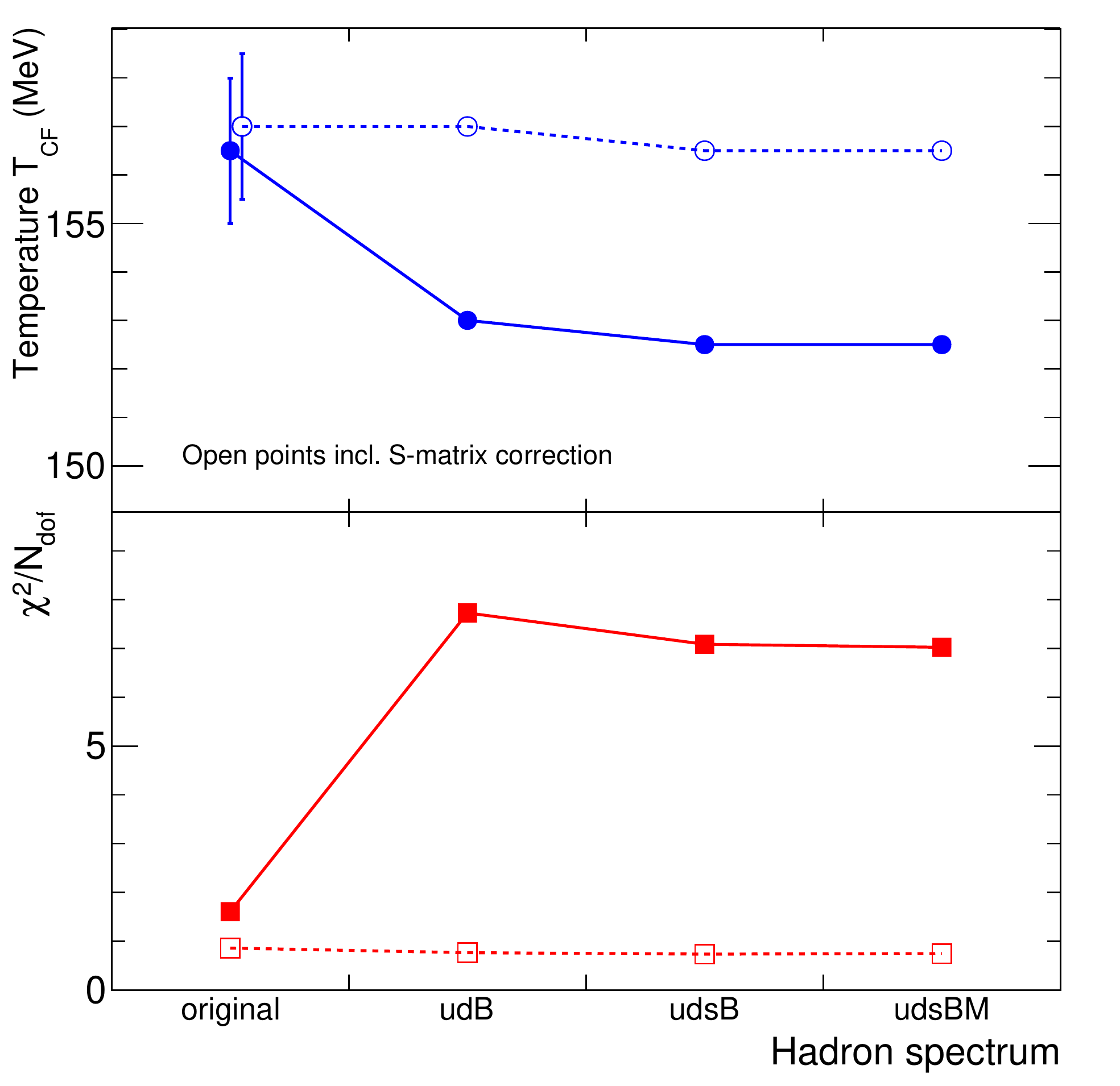}
    \caption{Chemical freeze-out temperature $\TCF$ (upper panel) and
      $\chisq$ (lower panel) for the 4 different model fits (solid
      symbols). The spectrum used is indicated as the x-axis
      label. S-matrix corrected values are shown as open points. The
      error bars indicate the uncertainty of the fitted temperature
      for the original PDG spectrum.}
    \label{fig:plot_T_chi2}
\end{figure}

In order to illustrate this effect quantitatively, we show in
Fig.~\ref{fig:plot_T_chi2} the values of the chemical freeze-out
temperature $\TCF$ obtained from the four different fits, and the
corresponding $\chisq$ values. The freeze-out temperature values from
the thermal fits decrease from $156.5$~MeV for the original PDG hadron
spectrum by about $4$~MeV for the fits with modified mass spectra. The
fit uncertainty of 1.5 MeV is indicated for the original PDG spectrum
fit. For all spectra the fit uncertainty is strongly correlated and
the drop is significant. On the other hand, the reduced $\chi^2$
increases from $1.6$ to a value of about $8$ for the 'udB' mass
spectrum where $N^*$ and $\Delta^*$ resonances are added, to drop
again slightly to a value a bit above 7 when strange baryons and
mesons are also added with the 'udsB' and udsBM' mass spectra.

As can be seen in Fig.~\ref{fig:plot_AllStates}, there are three $N^*$
states which have masses significantly lower than the other baryons
and well below the lightest known isospin 1/2 excitation, the
$N^*(1440)$. Already the low mass of the Roper resonance as first
radial excitation of the nucleon together with its electromagnetic
form factor has been a long standing puzzle only resolved
recently~\cite{Burkert:2017djo}. It is all the more surprizing to find
significantly lower isospin 1/2 excitations with $J^{\pi}$ = $1/2^-$
and $3/2^-$. We also note here that, in the baryon spectrum from the
constituent quark model~\cite{Capstick:1986bm}, such low-lying nucleon
resonances do not arise. As a cross check, these three states were
removed to evaluate their impact on the thermal fits. Indeed the
effect on the fit result and quality is significant: the fitted
temperature rises by 2 MeV and the $\chisq$ drops from 7.15 to 4.3.

Concurrently to the present study, an approach, which takes into
account the pion-nucleon interaction through the $S$-matrix
formalism~\cite{Lo:2017lym} was applied successfully to properly
reproduce the measured proton abundance at LHC energy using the
original PDG hadron spectrum~\cite{Andronic:2018qqt}. This approach
resulted in a moderate, temperature dependent reduction of the
computed proton yield, e.g. by about 17~\% at $\TCF$ = 156.5 MeV. This
brought the computed proton yield completely in line with the
experimental data, resulting in a thermal fit with a reduced $\chi^2$
value of 16.9/19, without change in the fit temperature. This approach
effectively resolved the only significant discrepancy between model
predictions and data, leading to an excellent description of all
experimental results.

To benefit from the recognition, in
refs.~\cite{Lo:2017lym,Andronic:2018qqt}, that taking into account via
the $S$-matrix approach non-resonant and repulsive components in the
pion-nucleon interaction for some partial waves leads to a much
improved description of the SHM proton yields, this approach is also
tested for the current modified hadron spectra with additional states.
We apply during the fitting procedure with each of the 4 hadron
spectra the appropriate correction to the proton yield as computed
in~\cite{Andronic:2018qqt}. The results are also displayed in
Fig.~\ref{fig:plot_T_chi2} by the open symbols. As reported
in~\cite{Andronic:2018qqt}, for the original PDG hadron spectrum the
resulting fit temperature is practically unchanged but the reduced
$\chi^2$ value drops from 29/19 $\approx$ 1.5 to 16.9/19 $\approx$
0.9. Also for the modified spectra with the additional states, this
results in a dramatic improvement compared to fits without this
$S$-matrix correction. Comparing the fit based on the PDG spectrum to
the one with the 180 additional nucleon excitations (spectrum 'udB'),
the fit result and quality are identical. The statistical model fit
yields identical results with or without the additional excited
nucleon states.

That implies that, after properly taking into account the pion-nucleon
interaction via the measured phase shifts and the according $S$-matrix
correction to the chiral susceptibilities and the pressure, there is
obviously no room for additional independent resonant
contributions. It is just as if these states did not exist. They leave
no trace in the phase shifts and virial expansion of the interaction
contribution to the pressure in the hadron resonance gas
model. Inspecting the phase shifts for the partial waves corresponding
to the low lying additional $N^*$ resonances that have the largest
effect on the proton yields computed without $S$-matrix correction,
i.e. the S11 and D13 for the $J^{\pi} = 1/2^-$ and $3/2^-$ states
given in~\cite{Lo:2017lym}, there is no indication for any other
resonant contribution. The integrated contribution of both the S11 and
D13 partial waves to the chiral susceptibility is about 30 \% less
than the naive resonance contribution for the PDG
states~\cite{Lo:2017lym} and consequently even less for a spectrum
with additional states of these quantum numbers. Similar conclusions
can be drawn for the other additional states vis-a-vis the partial
wave analysis. For the partial waves corresponding to $\Delta \,
1/2^-, 1/2^+, 3/2^-, 5/2^-, 5/2^+$ and the $N^* \, 3/2^+$ the
contribution to the chiral susceptibility is negative, corresponding
to repulsion in the respective $\pi N$ channel. This is not to say
these additional states do not exist, they may well. The phase shift
analysis and S-wave correction to the hadron re\-so\-nance gas (SHM)
tell us that they should not be treated as independent
resonances. When coupled to the continuum they could strongly
rearrange, inducing non-resonant or repulsive
interactions~\cite{Pok2020}.

For the spectra with additional strange baryons and additional mesons
the fits with $S$-matrix correction for the pion-nucleon interaction
are even slightly better as can be seen in
Fig.~\ref{fig:plot_T_chi2}. The currently available S-matrix
correction does not affect the yields of strange baryons and we
currently have no information on the effect of the pion-hyperon
interaction on the SHM predictions.

\section{Summary and conclusion}

We presented results of thermal fits to hadron yields measured at the
LHC in central Pb--Pb collisions, performed in the statistical
hadronization approach including additional baryon and meson states
proposed by LQCD calculations and a relativistic constituent quark
model. Adding these additional states to those obtained from the PDG
compilation leads to a strong deterioration of the thermal fit quality
as well as to a reduction of the chemical freeze-out temperature by
about 4 MeV. The deterioration is driven entirely by the added
non-strange baryons, the $N^*$ and $\Delta^*$ states. The application
of the $S$-matrix correction completely restores the quality of the
SHM fit with resulting temperature very close to 156.5 MeV.

These results indicate that the additional excited nucleon states, if
they exist, are not realized in nature as independent resonances. In
any case, the $S$-matrix treatment of the pion-nucleon interaction
contains the complete physical information and should be appropriate
for any spectrum as long as all of the corresponding partial waves for
the quantum numbers of the states used are included. The outcome of
this study is twofold. The added excited nucleon states $N^*$ and
$\Delta^*$ do not leave any trace in the thermal fits of the LHC data
after $S$-matrix correction for the pion-nucleon interaction. The
addition of a large number of excited hyperons and of some non-strange
and strange mesons gives slightly improved fits to experimental
yields. The effect on the extracted chemical freeze-out temperature is
negligible. Both observations show that the extraction of chemical
freeze-out parameters from the first moments of hadron multiplicities,
the event averaged yields, is rather robust against addition of even a
large number of hadronic states. Whether this is the case also for SHM
analysis of higher moments is currently a matter of
debate~\cite{Alba:2015iva,Bellwied:2017uat,Braun-Munzinger:2020jbk}.

\section{Acknowledgement}
We would like to thank Pok Man Lo and Krzysztof Redlich for
enlightening discussions. This work was funded in part by the DFG
(German Research Foundation) -- Project-ID 273811115 --
SFB1225/ISOQUANT.

\bibliographystyle{utphys}
\bibliography{references}

\end{document}